\documentstyle[12pt]{article}
\newcommand{\be}{\begin{equation}}
\newcommand{\im}{{\cal I}m}
\newcommand{\re}{{\cal R}e}
\newcommand{\ee}{\end{equation}}
\newcommand{\bea}{\begin{eqnarray}}
\newcommand{\eea}{\end{eqnarray}}

\newcommand{\pp}{~~~.}
\newcommand{\vv}{~~~,}
\newcommand{\slp}{/\!\!\!p}

\newcommand{\gapproxeq}
{\lower .7ex\hbox{$\;\stackrel{\textstyle >}{\sim}\;$}}
\newcommand{\lapproxeq}
{\lower .7ex\hbox{$\;\stackrel{\textstyle <}{\sim}\;$}}

\begin{document}
\setlength{\unitlength}{1mm}

{\hfill hep-ph/9912471}

{\hfill DSF 41/99}

\begin{center}
{\Large \bf Unstable Heavy Majorana Neutrinos and Leptogenesis}
\end{center}

\bigskip\bigskip

\begin{center}
{\bf Gianpiero Mangano} and {\bf Gennaro Miele}
\end{center}

\vspace{.5cm}

\noindent
{\it Dipartimento di Fisica, Universit\'{a} di Napoli "Federico II", and INFN,
Sezione di Napoli, Mostra D'Oltremare Pad. 20, I-80125 Napoli, Italy}\\

\begin{abstract}
We propose a new mechanism producing a non--vanishing lepton number
asymmetry, based on decays of heavy Majorana neutrinos. If they are
produced out of equilibrium, as occurs in preheating scenario, and are
superpositions of mass eigenstates rapidly decaying, their decay rates
contains interference terms provided the mass differences $\Delta m$ are
small compared to widths $\Gamma$. The resulting lepton asymmetry, which is
the analogue of the time--integrated $CP$ asymmetry in $B^0-\overline{B}^0$
system, is found to be proportional to $\Delta m/\Gamma$.
\end{abstract}
\vspace*{2cm}

\begin{center}
{\it PACS number(s): 98.80.Cq; 98.80.-k; 14.60.P}
\end{center}

\newpage
\baselineskip=.8cm

\section{Introduction}

A possible mechanism leading to the production, in the early universe, of a
baryon--antibaryon asymmetry can be found in terms of lepton number
production by heavy Majorana neutrino decays \cite{FukugitaYanagida}. As
well known, the lepton number $L_0$ so generated is reprocessed by
sphaleron transitions, and partially converted into baryon number B
\be
B = -\frac{8 n_g + 4 n_H}{22 n_g + 13 n_H}L_0~~~,
\label{1}
\ee
where $n_g$ is the number of fermion generations, and $n_H$ denotes the
number of Higgs doublets of electroweak Standard Model. There are two
crucial issues in this mechanism which eventually determine its efficiency
in producing the value for the baryon to photon number $\eta \sim
\left(10^{-10} \div 10^{-9}\right)$, fixed by observations on light
nuclei abundances.

First of all, the number density of heavy right--handed Majorana neutrinos
$N_i$ depends on the mechanism of reheating. Since the produced lepton
number is proportional to the number of $N_i$ per comoving volume, it is
quite important to have a prediction for this parameter, as function of
neutrino masses. This has been considered by many authors in the usual
scenario of neutrino production {\it via} thermal excitations in the bath
after reheating. In this case neutrino masses should be smaller than the
maximal temperature obtained during the reheating of the universe. It has
been recently shown by following in detail the reheating mechanism
\cite{Kolb} that a reasonable estimate for this upper bound is of the order
of $10^3 ~T_{RH}$, where $T_{RH} \simeq
\sqrt{\Gamma~ m_{Pl}}$,  is the so--called reheating temperature,
$\Gamma$ being the decay rate of the
inflaton.
The value of $T_{RH}$
is constrained to be in the range $\left(10^{8}\div
10^{10}\right)GeV$ in order to avoid an overproduction of gravitinos in
SUSY scenarios \cite{gravitino}, which would imply that very heavy neutrinos,
with masses larger than $10^{13}$ $GeV$ would not be significantly produced
in the reheating. Even stronger constraints on $T_{RH}$ have been obtained
by considering non thermal production of gravitinos \cite{gravitino2}.

More recently, the neutrino production has been also analyzed in the
so--called preheating scenario \cite{Giudiceetal}, which corresponds to a
resonant particle production during the first inflaton oscillations around
the minimum of the potential. The main result of this mechanism is that
even neutrinos with masses of the order of $10^{15}~GeV$ can be efficiently
produced in highly non--thermal momentum distribution. As will be clear in
the following, this feature of $N_i$ energy spectrum distribution is
crucial for our purposes.\\ The second issue is related to the presence of
$CP$ violating contributions to the neutrino decay channels in massless
fermions and Higgs bosons, and to the role of interference effects which
make this $CP$ violation observable. There are two such contributions to
the microscopic asymmetry $\epsilon$ which have been considered in
literature. The first one identified \cite{FukugitaYanagida} is due to
interference of the tree--level amplitudes with the absorptive part of the
one loop vertex ($\epsilon'$--like effect)
\be
\epsilon_v = - \frac{1}{8 \pi}\left[ \left(h^\dagger h\right)_{11} \right]^{-1}
\sum_{j} \im\left[\left(h^\dagger h \right)_{1j}^2\right]
f\left(\frac{m^2_j}{m^2_1} \right)~~~,
\label{2}
\ee
with $m_1$ the mass of the lightest Majorana neutrino, $f(x)$ can be found
in Ref. \cite{FukugitaYanagida} and $h$ is the coupling of $N_i$ to
massless left--handed leptons and Higgs bosons.

Recently, it has been observed that one should also consider contributions
coming via interference with one loop self--energy ($\epsilon$--like
effect) \cite{Roulet}-\cite{Pilaftsis}. For example for each neutrino $N_i$
one has
\be
\epsilon_s^i =- \frac{1}{8 \pi} \sum_{j\neq i}\frac{m_i m_j}{m_i^2-m_j^2}
\frac{\im \left[ \left(
h^\dagger  h\right)_{ij}^2\right]}{\left(h^\dagger  h\right)_{ii}}~~~.
\label{3}
\ee
It is worth observing that both asymmetries (\ref{2}) and (\ref{3})
very much resemble the $CP$ violation asymmetries for charged $B$ mesons
\be
\epsilon_f = \frac{\Gamma(B^+ \rightarrow f)-\Gamma(B^- \rightarrow f^c)}
{\Gamma(B^+ \rightarrow f)+\Gamma(B^- \rightarrow f^c)}\vv
\label{3bis}
\ee
with $f$ denoting an arbitrary final state. As for $\epsilon_v$ and
$\epsilon_s$, in order to have a not vanishing result one needs more than
one contribution to the exclusive decay channel $f$. In complete analogy to
lepton asymmetries, the additional contribution to exclusive $B^{\pm}$ decay
channels is provided by radiative processes, the so--called {\it penguin}
diagrams.

In this paper we consider a different scenario for neutrino decays, which
provides
already at tree--level in the amplitudes a microscopic contribution to the
lepton
asymmetry. The mechanism, following again the analogy with $B$ physics, is
much reminding the
time integrated $CP$ asymmetry in the $B^0-\overline{B}^0$ system
\cite{bsystem}. In this case, differently from the charged $B$, by virtue
of $B^0-\overline{B}^0$ oscillations one can produce
$CP$ asymmetries already at tree--level.

We consider the case in which heavy Majorana neutrinos are produced out of
thermal equilibrium through a preheating mechanism. For arbitrary couplings
of these neutrinos to the inflaton field, they are produced as
superpositions of the {\it mass eigenstates} $N_i$, that we denote as {\it
inflaton eigenstates}, $N_\alpha$. If neutrino lifetimes are less than the
typical decoherence time due to scattering processes in the medium, these
states, once produced, propagate as coherent superpositions of the $N_i$
till their eventual decay, if neutrino lifetimes are less than the typical
decoherence time due to scatterings in the medium. This constraint leads to
conditions on both the couplings $h$ and the Majorana masses which,
however, are neither particularly severe nor fine tuned.

As in the usual scenario, lepton number is produced via decays in
(massless) left--handed fermions $\psi_{Lj}$ ($j$ being the family index)
and Higgs bosons $\Phi$, and their $C$--conjugate particles, right--handed
antifermions $\psi_{Rj}^{c}$ and $\Phi^c$, giving rise to a microscopic
asymmetry, for each $\alpha$
\begin{equation}
\epsilon_\alpha = \sum_{j=1}^{n_g} \frac{\Gamma(N_\alpha \rightarrow
\psi_{Lj} \Phi) -
\Gamma(N_\alpha \rightarrow \psi_{Rj}^{c} \Phi^c)}{\Gamma(N_\alpha
\rightarrow \psi_{Lj} \Phi) +
\Gamma(N_\alpha \rightarrow \psi_{Rj}^{c} \Phi^c)}~~~.
\label{4}
\end{equation}
Since each $N_\alpha$ is a quantum superposition of the mass eigenstates
$N_i$, their decay amplitudes are linear combinations of ${\cal A}(N_i
\rightarrow \psi_{Lj} \Phi, ~\psi_{Rj}^{c} \Phi^c)$.
Provided the Yukawa matrix explicitly breaks $CP$ invariance, a not
vanishing value for $\epsilon_\alpha$ can be obtained at tree--level only
if these different amplitudes may interfere.

It should be stressed at this point that it is always difficult to deal
with unstable particles in the clean framework of Quantum Field Theory,
since, in this case, they cannot be identified with asymptotic states in
some Hilbert space, but rather they appear as resonances in S--matrix
elements. An evaluation of the asymmetries $\epsilon_v$ and $\epsilon_s$ in
this approach has been done in \cite{Frere}. In our case, one should
consider the decay of the inflaton field into the states $\psi_{Lj}
\Phi$, $\psi_{Rj}^{c}
\Phi^c$ which resonantly proceeds via intermediate $N_\alpha$.
If, however, the heavy neutrinos have very small decay widths compared with
their masses, which we will always assume in the following, one may
envisage this process as the production of the (quasi)-stable state
$N_\alpha$ and its subsequent decay. To describe this process we will
introduce a neutrino wave function, accounting for its exponential decay,
whose espression can be deduced from the form of the corresponding
propagator in the vicinity of the resonance.

The basic idea of the paper is as follows. Let us consider an arbitrary
final state, like $\psi_{Lj}
\Phi$ or $\psi_{Rj}^{c}
\Phi^c$, with invariant mass $\mu$. It can be produced by the decay of
neutrinos with mass $m_i$ if $\mu$ is, say, in the range $m_i-\Gamma_i/2\leq
\mu \leq m_i+\Gamma_i/2$. If there are two such neutrinos $N_i$ and $N_k$,
whose mass difference is smaller than their $average$ width
$\Gamma_{ik}=(\Gamma_{i}+\Gamma_k)/2$, it is impossible to distinguish if
this decay product is the result of the decay of $N_i$ or rather $N_k$. In
other words, for the process $N_\alpha \rightarrow \psi_{Lj}
\Phi$ one expects in this case the contribution of the amplitudes due to
both possible processes: $ N_\alpha
\rightarrow N_i \rightarrow \psi_{Lj} \Phi$ and $N_\alpha \rightarrow N_k
\rightarrow \psi_{Lj} \Phi$.\\
The interference between the two amplitudes leads, in general, to not
vanishing microscopic asymmetries $\epsilon_\alpha$, which are found to be
proportional to the factors $(m_i-m_j)/\Gamma_{ij}$ and to the imaginary
parts of the matrix products $h h^\dagger$. Of course, if the Majorana
neutrinos would be absolutely stable states, or, in the case of decaying
particles, if their mass differences are much greater than their widths
$\Gamma_i$, any interference would be impossible and the resulting
asymmetry would vanish.

The paper is organized as follows. In section 2 we briefly summarize the
main features of the model describing heavy Majorana neutrino dynamics,
whose out of equilibrium production mechanism is discussed in section 3.
The estimate of the resulting macroscopic $L$ asymmetry is presented in
section 4. Finally, section 5 contains our conclusions and outlooks.

\section{The model}

Let us consider heavy neutrinos $\nu_{R i}$ and $\nu^c_{L i}$ ($i$
denotes the family index) with a Majorana mass term
\be
L_M = -  \left(\overline{\nu}^c_{L i} \, M_{i j}
\, \nu_{R j}\, + \, \overline{\nu}_{R i} \, M_{i j}
\, \nu^c_{L j}\right)\vv \label{5}
\ee
with $M$ a symmetric real matrix. This mass term can be diagonalized in
terms of a set of Majorana neutrinos $N_i$, with a latin letter as family
index, defined as follows
\be
\nu_{Ri}  = P_R \, W_{ij} \, N_j \vv ~~~
\nu^c_{Li} = P_L \, W_{ij} \, N_j \vv \label{8}
\ee
where $W$ is an orthogonal matrix such that $W^T M W$ is diagonal, and $P_{R,L}
\equiv (1{\pm} \gamma_5)/2$. Denoting with $\chi$ the scalar field behaving,
in a certain period of the evolution of the universe, as the inflaton, the
production of the heavy neutrinos via the reheating mechanism takes place
due to a Yukawa term in the Lagrangian density of the form
\be
L_\chi = - \chi \,\left( \overline{\nu}^c_{Li} \, G_{ij}
\, \nu_{Rj} \, + \, \overline{\nu}_{Ri} \, G_{ij}
\, \nu^c_{Lj}\right)\vv
\label{9}
\ee
where again $G$ is a symmetric real matrix. Let us denote with $N_{\alpha}$
(with a greek letter as index) the basis of $G$ eigenstates. In this basis
the Majorana mass matrix $M$ is in general not diagonal, the two basis
being connected by an orthogonal transformation
\be
N_\alpha = U_{\alpha i} \, N_{i}\pp
\label{10}
\ee
Finally, the heavy neutrinos are also coupled to massless left--handed
leptons and to the standard $SU(2)$ Higgs doublet $\Phi$ through a Dirac
term
\be
L_D = - \, \left(\overline{\psi}^c_{Ri}{\cdot}\Phi \right)h^{*}_{ij} N_j -
\, \left(\overline{\psi}_{Li}{\cdot}\Phi^c \right) h_{ij} N_j \vv
\label{11}
\ee
with $h$ the Yukawa coupling matrix in the family space and
\be
\psi_{Li} = \left( \begin{array}{c}
\nu_{Li} \\ \\ l^-_{Li}\end{array}\right)
\, \vv ~~~~~~~~~~~~~~~~\Phi = \left( \begin{array}{c} \varphi^+
\\ \\ \varphi^0   \end{array}\right) \pp
\label{a12}
\ee

\section{Neutrino production}

In the usual scenario, neutrinos, along with all species of particles with
masses below the maximal temperature achieved during reheating,
are produced as thermal excitations
when the inflaton releases its energy density and the radiation epoch
starts. This means that, if the $N_i$ have masses smaller than this temperature,
they can be thermally excited. In this case, the $N_i$ field configuration
would correspond to a thermal distribution of particles with definite mass
given by the eigenvalues $m_i$ of the Majorana mass matrix $M$. When
eventually the temperature decreases down to $m_i$, if the decays into
massless leptons and Higgses take place in a out--of--equilibrium
condition, then a macroscopic lepton asymmetry can be produced, provided
all Sacharov conditions are satisfied. This scenario has been widely
studied, and, as mentioned in the previous section, the effect is basically
due to the interference between the tree--level decay amplitudes and the
one--loop contributions \cite{FukugitaYanagida,Roulet}. In light of the bounds
on $T_{RH}$ and the discussion in the Introduction, this scenario seems only
viable for Majorana neutrinos with masses lighter than $10^{13}$ $GeV$.

We would rather analyze the case when these neutrinos are produced through
a preheating mechanism as coherent superpositions of mass eigenstates. In
this scenario particles are resonantly excited due to the oscillatory
behaviour of the inflaton field $\chi(t) = \chi_0 \, cos(m_\chi t)$,
$m_\chi^2$ being the second derivative of the inflaton potential at its
minimum. In the following we will consider the simplest case of a quadratic
potential for $\chi$. It has been shown by several authors,
\cite{gravitino2}, \cite{Giudiceetal}, \cite{linde1}--\cite{anupam}, that
is possible to produce spin$(0)$, spin$(1/2)$, as well as spin$(3/2)$
particles through non--perturbative effects. The neutrino field satisfies a
Dirac equation on a Friedmann--Robertson--Walker spacetime with an
effective time dependent mass matrix ${\cal M}$. In conformal time $d\eta =
dt/R(t)$, with $R(t)$ the scale factor
\begin{equation}
\label{eqn}
\left[\frac{i}{R}\gamma^{\mu}\partial_{\mu}+i\frac{3}{2 R^2}
\frac{d R}{d\eta}\gamma^{0}\right] N_i = {\cal M}_{ij}(\eta) N_j \vv
\label{a13}
\end{equation}
with
\be
{\cal M}_{ij}(\eta) = M_{ij} + \chi(\eta) \,G_{ij} ~~~.
\label{a14}
\ee
Equation (\ref{a13}) describes oscillators with a complex time varying
frequency. In the simple case of diagonal $M$ and $G$ matrices the neutrino
number density is obtained by using a time dependent Bogoliubov canonical
transformation \cite{Giudiceetal}. In the general case, however, since $M$
and $G$ are not necessarily simultaneously diagonal one should first
diagonalize via an orthogonal transformation the effective mass ${\cal M}$,
whose eigenvalues may have quite an involved dependence on $\chi(\eta)$.
Let us consider the two possible cases:
\begin{itemize}
\item[i)] if the order of magnitude of the matrix elements of $M$ is much larger
than the one of $\chi(\eta) \,G $, we may treat the term $\chi(\eta) \,G $
as a perturbation to $M$. Thus in the Majorana mass eigenbasis,
up to the first order in perturbation theory we get
\begin{equation}
{\cal M}_{ij}(\eta) \simeq \left[m_i + \chi(\eta)\, G_{ii} \right]
~\delta_{ij}
\pp
\label{a15}
\end{equation}
Hence the $n_g$ equations for $i=1,2,..,n_g$ are decoupled and can be
treated as in Ref. \cite{Giudiceetal}. In this case the inflaton would
produce neutrinos already as mass eigenstates, but the resonant condition
$\mbox{det}\left[{\cal M(\eta)}\right]=0$ for an explosive production of
heavy neutrinos is never satisfied.
\item[ii)] in the opposite case, when the time dependent mass term,
provided by the coupling to the inflaton, is comparable or even dominant over
$M$, it is convenient to use the {\it inflaton} basis to write the mass term
as
\begin{equation}
{\cal M}_{\alpha \beta}(\eta) \simeq \left[M_{\alpha \beta} +
\chi(\eta) \, g_\alpha \delta_{\alpha \beta}  \right]~
~~~.
\label{a16}
\end{equation}
Thus, the preheating neutrino production occurs for those values of $\eta$
for which $\mbox{det}\left[{\cal M}(\eta)\right]$ vanishes.
\end{itemize}

In case $ii)$, if we assume, as it will be clear in the following, small
off--diagonal terms in the matrix $M$ compared to the diagonal ones, we can
write the preheating production condition as
\be
\mbox{det}\left[{\cal M}(\eta)\right] = \Pi_{\alpha=1}^{n_g}
\left( M_{\alpha \alpha} + \chi(\eta) \, g_\alpha \right)+
O\left(\frac{m_{\alpha \beta}}{\overline{m}} \right) = 0 \vv
\label{a16a}
\ee
where $m_{\alpha \beta}$ just denotes the order of magnitude of the
off--diagonal mass terms in $M$ and $\overline{m}$ the one of the diagonal
entries. At lowest order in the ratio $m_{\alpha \beta}/\overline{m}$, eq.
(\ref{a16a}) is satisfied if, for some $\eta_*$
\be
M_{\alpha \alpha} + \chi(\eta_*) \, g_\alpha =0 \pp
\label{a16aa}
\ee
In this case $N_\alpha$ heavy Majorana neutrinos will be resonantly
produced. However, since the conditions (\ref{a16a}) is only satisfied up
to terms of the order $m_{\alpha \beta}/\overline{m}$, this implies that
the production rates are suppressed to some extent. In other words it is as
these neutrinos at $\eta_*$ would not be produced as massless but with a
mass of the order of $m_{\alpha \beta}^2/\overline{m}$. This effect which
leads to an exponential suppression factor of the number of heavy Majorana
neutrinos produced \cite{Giudiceetal}, can be neglected if the following
condition is satisfied
\be
\frac{m_{\alpha \beta}^2}{\overline{m}} << \sqrt{g_\alpha ~\chi'(\eta_*)}\pp
\label{a16b}
\ee
Since $g_\alpha \chi'(\eta_*) = g_\alpha m_\chi \chi(\eta_*)
\sim \overline{m} ~m_\chi$,
the above condition becomes
\be
\frac{m_{\alpha \beta}}{\overline{m}} << \left(\frac{m_\chi}{\overline{m}}
\right)^{1/4}
\pp
\label{a16c}
\ee
The constraint (\ref{a16c})
gives for example for the typical values
$m_\chi=10^{13}~GeV$ and $\overline{m}=10^{15}~GeV$, $m_{\alpha \beta} /
\overline{m} \leq 10^{-1}$, which does not severely affects
the order of magnitude of the off--diagonal elements in $M_{\alpha \beta}$,
and still allows for a quite large mixing, given by the non diagonal
elements of the matrix $U$, see eq. (\ref{10}).

Under the condition (\ref{a16c}) one can safely apply the results of Ref.
\cite{Giudiceetal} where to solve eq. (\ref{a13}), in case $ii)$,
one writes down the momentum expansion for Majorana neutrinos
$N_\alpha(\eta,\vec{x})$.
By a time dependent Bogoliubov transformation
it is then possible to diagonalize the Hamiltonian in terms of
quasi--particle creation and annihilation operators and, with a customary
procedure, to deduce the number of produced particles as the expectation value
of the particle number operator $n_\alpha$ on the quasi--particle vacuum.
Depending on the value
of the parameter $q_{\alpha}$, defined as
\be
q_\alpha \equiv \frac{g_{\alpha}^2 ~\chi^2 (0)}{4 ~m_\chi^2} \vv
\label{11-9}
\ee
with $\chi(0)$ the initial value of the inflaton field configuration, the
final number density quite rapidly reaches the bound due to Pauli blocking
and can be expressed in terms of the maximal momentum $k_{max}$ of the
distribution
\be
n_\alpha \simeq k_{max}^3 \simeq m_\chi^2 M_{\alpha \alpha} \pp
\label{11-10}
\ee
At later times, when eventually oscillations are damped to smaller values,
the fraction of the inflaton energy density transferred to heavy neutrinos,
$\rho_\alpha/\rho_\chi$, is frozen to the value, see Ref.
\cite{Giudiceetal},
\be
\frac{\rho_\alpha}{\rho_\chi} \simeq \frac{ m_\chi^2}{\chi^2(0)}~ q_\alpha \pp
\label{11-11}
\ee
In the framework of a chaotic inflation scenario, from the observed
amplitude of the density perturbations on large scales, $m_\chi$ is
constrained to be of the order of $10^{13} ~GeV$. Furthermore,
with $\chi(0)\simeq
m_{Pl}$, one gets $\rho_\alpha
/\rho_\chi \simeq 10^{-12} q_\alpha$. In the following this result will be
used to estimate the final lepton number produced by heavy neutrinos.

\section{Neutrino propagation and decay}

In the previous section we have described the production mechanism of the
$N_\alpha$ neutrinos regarded as asymptotic stable states. In principle the
neutrino production by the inflaton, and their subsequent decays in
fermion-Higgs pairs should be considered as a whole, $\chi \rightarrow
N_\alpha \rightarrow \psi_{Lj} \Phi, \psi_{Rj}^{c} \Phi^c$. The metastable
character of the $N_\alpha$ would therefore be encoded in the expression of
its propagator. However it should be first noted that the nonperturbative
mechanism outlined in the previous section is basically istantaneous
\cite{Giudiceetal}. If in addition the $N_\alpha$ are sufficiently
long-lived, i.e. the widths are small compared with their masses, we can
safely describe the process as occurring in three stages: the production of
the quasi-stable states $N_\alpha$, its propagation in the medium and its
eventual decay.

If neutrinos are produced in the inflaton basis $N_\alpha$, say at time
$t=0$ and with momentum $\vec{k}$, they start evolving as a linear
superposition of the mass eigenstates $N_j$ of eigenvalue $m_j$ as follows
\be
N_\alpha(x;\vec{k}) = U_{\alpha j} \, N_j(x;\vec{k})~~~,
\label{15}
\ee
with $x\equiv (t,\vec{x})$ and $N_j(x;\vec{k})$ the wave-function of the
mass eigenstate $N_j$. The form of this wavefunction, which account for the
metastable behaviour of $N_j$ and can be inferred from the form of the
propagator, will be discussed in the following. In presence of a medium,
scattering processes will tend to destroy the coherence among the
components of the wave function $N_\alpha(x;\vec{k})$ before its decay, and
to populate the universe with thermalized neutrino mass eigenstates. In
order to avoid this, one has to impose the condition
\be
\Gamma_\alpha >> n \langle \sigma \rangle  \equiv \Gamma_{sc}~~~,
\label{15bis}
\ee
where $\Gamma_\alpha$ denotes the $N_\alpha$ decay rate
\be
\Gamma_\alpha =  U_{\alpha i}  \left(
\sum_f \langle f | N_i
\rangle  \langle N_j | f \rangle \right) U_{j \alpha}^T ~~~,
\label{16}
\ee
where the sum is over all possible final states $f$. With $\sigma$ we
denote the cross section of the relevant scattering processes, averaged
over the incoming particles distribution with number density $n$. The
dominant contribution to $\Gamma_\alpha$ comes from the two body decay
channels $\psi_{Li} \Phi $. If the neutrino masses are of the same order of
magnitude $m_i \simeq \overline{m}$ $\forall i$, the factor in bracket in eq.
(\ref{16}) simplifies to
\be
\sum_f \langle f | N_i
\rangle \langle N_j | f \rangle \simeq (h h^\dagger)_{ij} ~
\frac{\overline{m}}{8 \pi} \Rightarrow
\Gamma_\alpha \simeq (h h^\dagger)_{\alpha \alpha} ~ \frac{\overline{m}}
{8 \pi}\pp
\label{16bis}
\ee
Notice that the case of almost mass degenerate neutrinos is actually the
scenario we are mostly interested in this paper.

The main contributions to
$\Gamma_{sc}$ correspond to the processes $N_\alpha \, \Phi
\longrightarrow N_i \, \Phi$, $N_\alpha \, \psi_{L i}
\longrightarrow N_i\, \psi_{L j}$ and crossed channels, as well as to
scattering over the large amount of inflaton quanta $\chi$. Condition
(\ref{16}) constraints more severely the Yukawa couplings $h$ and $G$ when
the mean energy of the massless $\Phi$, $\psi_{L i}$ and $\chi$ is larger
than the neutrino masses. This is due to the rapid increasing with this
mean energy of the product $n_{\Phi,\psi_L,\chi} \, \langle \sigma \rangle
$. In fact, defining, in the $N_\alpha$ rest frame, the {\it effective}
temperature $T_*$, which represents the mean energy of the massless
fermions or Higgses, and similarly $T_{\chi}$ the one of the inflaton
excitations, we have, for scattering over massless fermions and Higgses
\be
\langle \sigma(\psi_L,\Phi) \rangle \simeq \frac 1{8 \pi ~\overline{m} ~T_*}
\left[(h h^\dagger)_{\alpha \alpha}
\sum_{j=1}^{n_g} (h h^\dagger)_{jj} + ( h h^\dagger h h^\dagger)_{
\alpha \alpha}\right]
\vv
\label{17}
\ee
and similarly, for scattering processes over $\chi$ bosons
\be
\langle \sigma(\chi) \rangle \sim \frac{1}{8\pi~
\overline{m} ~T_\chi} ~(G^4)_{\alpha \alpha}\pp
\label{17bis}
\ee
For low momentum neutrinos $T_*$ and $T_{\chi}$ also give the mean energy
in the comoving frame. Actually from the discussion of section 3, we see
that in our preferred choice for $\overline{m} \geq 10^{13}$ $GeV$ and
$m_{\chi} \sim 10^{13}$ $GeV$, neutrinos are basically emitted as
non--relativistic particles, since $ k_{max}/\overline{m} \simeq (m_\chi
/\overline{m})^{2/3} \leq 1$, thus we can safely neglect any effect due to the
difference between the neutrino rest frame and the comoving frame. Since
the number density of incoming particles can be expressed as
$n_{\psi_L}\simeq n_{\Phi} \simeq T_*^3$, $n_{\chi} \simeq T_\chi^3$, the
condition for no decoherence becomes
\be
{(h h^\dagger)_{\alpha \alpha}
\sum_{j=1}^{n_g} (h h^\dagger)_{jj} + ( h h^\dagger h h^\dagger)_{
\alpha \alpha} \over (h h^\dagger)_{\alpha \alpha}} T_*^2 +
\frac{(G^4)_{\alpha \alpha}}{(h h^\dagger)_{\alpha \alpha}} T_\chi^2 \leq
\overline{m}^2~~~.
\label{18}
\ee
No particularly fine tuned condition on the Yukawa couplings $h$ and $G$
follows from (\ref{18}) even for very heavy Majorana neutrinos,
$\overline{m} \sim 10^{15}~ GeV$. Assuming in fact the extreme limit
$T_\chi \sim \left( 10^{-2} \div 10^{-1}\right)~m_{Pl}$ which, still
compatible with a classical description of spacetime structure, seems to be
suggested by an exponential production of $\chi$ quanta by the inflaton
oscillating configuration \cite{explinde}, we get, as an extremely
conservative estimate, that there is no decoherence if we take the Yukawa
couplings $G$, $h$ up to $\left(10^{-3} \div 10^{-2}\right)$. This implies
the {\it conservative} upper bound for $q_\alpha$ (\ref{11-9})
\be
q_\alpha \lapproxeq \left( 10^{6} \div 10^{8}\right) \pp
\label{18bis}
\ee
If condition (\ref{18}) is satisfied, heavy neutrino states (\ref{15})
evolve coherently till their decays into massless fermions and Higgses.

To evaluate the decay rates in these channels we first have to specify the
expression for the wave functions $N_j(x;\vec{k})$. For this purpose we
first recall that, in a neighborhood of the mass eigenvalues $m_j$, the
$N_j$ propagator $G_j(p^2)$ gets the familiar Breit-Wigner behaviour. Using
Lehman spectral decomposition we can also write $G_j(p^2)$ as the
superposition of free propagators of given masses via two spectral
functions
\be
G_j(p^2)=\frac{\slp + m_j}{p^2-m_j^2+ i m_j \Gamma_j}~ {\cal
C}^{-1}=\int_0^\infty ~d \mu^2 \left( \slp ~w^1_j(\mu^2) + w^2_j(\mu^2)
\right)
\frac{1}{p^2-\mu^2+ i \epsilon} ~{\cal C}^{-1}
\vv
\label{bw}
\ee
with ${\cal C}$ the charge conjugation operator. For $\mu^2 \sim m_j^2$, it
is straightforward to get
\be
w^1_j(\mu^2)=\frac{w^2_j(\mu^2)}{m_j} = \frac{m_j \Gamma_j}{\pi \left[
\left(\mu^2-m_j^2\right)^2+ m_j^2 \Gamma_j^2 \right]} ~~~\pp
\label{spectr}
\ee
This result follows by expanding the free propagators as the principal
value of $(p^2-\mu^2)^{-1}$ and its Dirac $\delta$--contribution.

The positive definite spectral function $w^1_j$ can be simply related to
the wave function $\rho_j$ of the unstable state $N_j$ in configuration
space. If we consider the expansion of the field ${\cal N}_j$, which
represents the Majorana neutrinos, as follows
\be
{\cal N}_j(x) = \int \frac{d^3 k}{2 k^0} \left[ N_j(x;\vec{k}) ~a_{\vec{k}}
+ N_j^c(x;\vec{k}) ~a_{\vec{k}}^\dagger \right]
\ee
with $N_j(x;\vec{k})$ a set of neutrino wavefunction with spatial momentum
$\vec{k}$, encoding its unstable character, it is straightforward to see
that we recover the correct Breit-Wigner behaviour of the corresponding
propagator if we represent these wavefunctions as
\be
N_j(x;\vec{k})= \int d \mu^2 \rho_j(\mu^2)~ u^{\mu}_{\vec{k}} ~e^{-i k {\cdot}
x}~~~\vv
\label{20}
\ee
with $|\rho_j(\mu^2)|^2 = w^1_j(\mu^2)$. We have denoted with
$u^{\mu}_{\vec{k}}(x)$ a four component spinor solution of the Dirac
equation in momentum space with mass $\mu$ and momentum $\vec{k}$. The
spectral function $w^1_j$ does not uniquely fix the wavefunctions
$N_j(x;\vec{k})$. The simplest choice for $\rho_j(\mu^2)$, satisfying
$|\rho_j(\mu^2)|^2=w^1_j(\mu^2)$, is again a Breit--Wigner function
\be
\rho_j(\mu^2) = \frac{\sqrt{m_j \Gamma_j}}{\sqrt{\pi}\left(\mu^2-m_j^2+
i m_j \Gamma_j\right)}~~~.
\label{21}
\ee
We stress that in what follows it is not really crucial the particular
choice for $\rho_j(\mu^2)$, satisfying $|\rho_j|^2=w^1_j$. The only
relevant aspect we have to require in order to get a non vanishing lepton
asymmetry is that $\rho_j(\mu^2)$ has an imaginary part depending on the
two parameters $m_j$ and $\Gamma_j$. This assumption is quite robust, if we
one recall the fact that phase shifts due to the production of any
resonance is strongly dependent on these quantities. Finally we observe
that in the limit of vanishing decay widths $\rho_j(\mu^2)
\rightarrow \delta(\mu^2 - m_j^2)$, so in this limit $N_j(x;\vec{k})
\rightarrow u^{m_j}_{\vec{k}}(x)  e^{-i k {\cdot} x}$, which describes a
stable neutrino with mass $m_j$ and momentum $\vec{k}$

We can now evaluate the decay rates of the $N_\alpha$. As we have discussed
in the preheating mechanism neutrinos are basically produced with low
momenta, much smaller than their masses. It is therefore a good
approximation to consider the case of neutrinos emitted at rest in the
comoving frame. From eq. (\ref{11-10}) all corrections to the results below
are at most of the order of $(\overline{m}/m_\chi)^{2/3}$.

Depending on the Majorana neutrino mass spectrum, it is possible that two
or more $\rho_j$ significantly overlap. This occurs whenever $|m_i-m_j|
<(\Gamma_i+\Gamma_j)/2$. Using (\ref{20}), the total decay rate of
$N_\alpha$ into  pairs $\psi_{Lp}
\Phi$ as well as into the $C$--conjugated channels $\psi_{Rp}^c \Phi^c$
is given by
\be
\sum_{p=1}^{n_g} \Gamma(N_\alpha \rightarrow \psi_{Lp} \Phi) =
\Xi_\alpha^{ij} \, I_{ij} \vv~~~~
\sum_{p=1}^{n_g}  \Gamma(N_\alpha \rightarrow \psi_{Rp}^c \Phi^c) =
\left(\Xi_\alpha^{ij}\right)^* \, I_{ij} \vv
\label{23}
\ee
where
\begin{eqnarray}
\Xi_\alpha^{ij} &=& U_{\alpha i} (h h^{\dagger})_{ij} U_{\alpha j}
=\left(\Xi_\alpha^{ji}\right)^* \vv~~~\mbox{no sum over $i$ and $j$}\vv
\label{25} \\
I_{ij} &=& \frac{1}{8 \pi}\int_0^\infty \frac{\mu^2 \,  d\mu}
{\sqrt{|\vec{k}|^2+\mu^2}}  \, \rho_i(\mu)
\,\rho^*_j(\mu)\,\sim \frac{1}{8 \pi}\int_0^\infty \sqrt{\mu^2} \,  d\mu
\, \rho_i(\mu) \,\rho^*_j(\mu)
 = I_{ji}^* \pp\label{26}
\end{eqnarray}
In eq. (\ref{26}) the factor $\mu^2/(8 \pi
\sqrt{|\vec{k}|^2+\mu^2})\sim \mu/(8 \pi)$ is the result of integration over the
phase space for final massless particles with fixed initial mass $\mu$ and
momentum $\vec{k}$. Notice also that we have integrated in $I_{ij}$ over
the range $[0,\infty[$, rather than in the narrow neighborhood of the
masses $m_i$ and $m_j$. This is justified by observing that $I_{ij}$
receive the main contribution from these neighborhoods, since the products
$\rho_i(\mu^2) \rho^*_j(\mu^2)$ are rapidly decreasing functions for $\mu^2
<< m_i^2,m_j^2$ and $\mu^2>>m_i^2,m_j^2$.

A straightforward computation shows that the microscopic asymmetries
$\epsilon_\alpha$ are given by
\be
\epsilon_\alpha=\frac{2 \sum_{i<j} \im\left[\Xi_\alpha^{ij}
\right] \, \im
\left[I_{ij} \right]}{2 \sum_{i<j} \re \left[
\Xi_\alpha^{ij} \right] \, \re
\left[I_{ji}\right]+ \sum_{i=1}^{n_g} \Xi_\alpha^{ii} \, I_{ii}} \pp
\label{24}
\ee
From this result we get that, in order to have at tree--level not vanishing
microscopic asymmetries it is necessary that:
\begin{itemize}
\item[i)] a $CP$ violating term is contained in the Yukawa couplings $h$;
\item[ii)] the integrals $I_{ij}$ of the spectral functions contain
a not vanishing imaginary part.
\end{itemize}
As already stated, this second condition is realized if at least two of the
neutrino masses satisfy the condition $|m_j-m_i| \leq
(\Gamma_j+\Gamma_i)/2$, otherwise the two kernels $\rho_i(\mu^2)$
$\rho_j(\mu^2)$ have no significant overlap. If $m_j>m_i$, the main
contribution to the imaginary part of $I_{ij}$ is expected for
$m_i<\mu<m_j$. In this interval the phase difference of $\rho_i$ and
$\rho_j$ is almost $\pi$. Furthermore, even if $m_i=m_j$, an imaginary part
for $I_{ij}$ is expected if the two widths are sensibly different.

Defining $\Gamma_{ij}\equiv(\Gamma_i+\Gamma_j)/2$, $m_{ij}
\equiv(m_i+m_j)/2$, $\Delta_{ij}\equiv (m_j-m_i)/(\Gamma_i+\Gamma_j)$ and
$\gamma_{ij}\equiv (\Gamma_j-\Gamma_i)/\Gamma_{ij}$, and if we also assume
for simplicity that both $\Delta_{ij}$, $\gamma_{ij}<1$, the asymmetry can
be obtained as an expansion in powers of these parameters. In the narrow
width limit, a simple calculation up to the first order in $\Delta_{ij}$
and $\gamma_{ij}$ gives
\be
I_{ij} \simeq \frac{1}{8 \Gamma_{ij} }\left(1+ i \Delta_{ij} + i
\gamma_{ij}
\frac{\Gamma_{ij}}{m_{ij}} \right)~~~.
\label{28}
\ee
It is interesting to consider the case of only two generations, for which
the expression of the $L$ microscopic asymmetry is particularly simple and
the orthogonal matrix $U$ is given by
\be
U= \left(
\begin{array}{cc}
\cos(\theta) &
\sin(\theta)
\\
-\sin(\theta) &
\cos(\theta)
\end{array}\right) \pp
\label{29}
\ee
A simple calculation gives at the lowest order in $\Delta_{12}$ and
$\gamma_{12}$
\bea
\epsilon_{1,2}& = & {\pm} \frac{ 2 \sin(2 \theta) \im \left[
(h h^\dagger)_{12}
\right]}
{(h h^\dagger)_{11}+(h h^\dagger)_{22} \mp  \sin(2 \theta) \re \left[ (h
h^\dagger)_{12}
\right]} \left( \Delta_{12} + \gamma_{12} \frac{\Gamma_{12}}{m_{12}} \right)
\nonumber \\
&= & {\pm}   \sin(2 \theta) \lambda_{CP}\left( \Delta_{12} + \gamma_{12}
\frac{\Gamma_{12}}{m_{12}} \right)
\frac{ (h h^\dagger)_{11}+(h
h^\dagger)_{22}} {(h h^\dagger)_{11}+ (h h^\dagger)_{22} \mp \sin(2 \theta)
\re \left[(h h^\dagger)_{12}\right]}
\vv
\label{30}
\eea
with
\be
\label{30bis}
\lambda_{CP} = \frac{ 2 \im \left[
(h h^\dagger)_{12}\right]}{(h h^\dagger)_{11}+ (h h^\dagger)_{22}} \vv
\ee
representing the strength of the $CP$ violating effects. Notice that since
we are working with Majorana neutrinos, even for two generations a phase in
the matrix $h$ cannot be washed away by a simple redefinition of fields, so
$(h h^\dagger)_{12}$ is in general a complex quantity.

There are several features of eq. (\ref{30}) which is worth observing.
First of all the two asymmetries tend to cancel each other, simply because
the two numerators of $\epsilon_1$ and $\epsilon_2$ are opposite by the
orthogonality of the mixing matrix $U$. This is exactly analogous to $GIM$
mechanism \cite{Okun}. However, already at first order the sum
$\epsilon_1+\epsilon_2$ does not vanish because the denominator in eq.
(\ref{30}) is different for the two neutrinos. Furthermore, the total
lepton asymmetry $L$ is given by the microscopic asymmetries weighted by
the corresponding neutrino number densities $n_i$. As long as $n_1 \neq
n_2$ the value of $L$ is not expected to vanish, though a partial
cancellation still takes place.  It should also be pointed out that at
lower order there is also a  contribution from the widths difference
parameter $\gamma_{12}$, which is however suppressed by the small factor
$\Gamma_{12}/m_{12}$.

Finally notice that the value for the asymmetry is obtained at tree--level
in decay amplitudes, while both contributions previously considered,
$\epsilon_v$ and $\epsilon_s$, depend on matrix elements of $(h
h^\dagger)^2$ in the numerator, see eq.s (\ref{2}),(\ref{3}). This means
that the asymmetries (\ref{30}) can be in principle quite large since they
are not suppressed by higher powers of the Yukawa couplings, though there
is a certain cancellation among them. Also notice that, differently than
eq. (\ref{3}), $\epsilon_{1,2}$ are only linearly dependent on
$\Delta_{12}$, so they are less suppressed in the limit of small
$\Delta_{12}$. Incidentally, in the usual scheme the Yukawa matrix elements
are constrained to be quite small $(h h^\dagger)_{ii} \leq m_i/m_{Pl}$ to
have an out--of--equilibrium decay. This bound is no more necessary if the
neutrino are already produced, as in the preheating scenario, in a non
thermal way, and the reheating temperature $T_{RH}$ is low enough to
prevent from a subsequent thermal production of $N_i$.

Using the results of section $3$, we can finally estimate the total lepton
number $n_L$, normalized to specific entropy. Since the energy fractions
$\rho_\alpha$ remain constant till the inflaton decay into radiation, after
the reheating stage one gets
\be
\frac{n_L}{s} \simeq \frac{T_{RH}}{\overline{m} ~\rho_\chi}
\sum_{\alpha=1}^{n_g} \epsilon_{\alpha} \, \rho_\alpha
\simeq 10^{-17}
\left( \frac{T_{RH}}{10^{10} GeV} \right)
 \left( \frac{10^{15} GeV}
{\overline{m}} \right) \sum_{\alpha=1}^{n_g} \epsilon_{\alpha} \, q_\alpha
\pp
\label{30ter}
\ee
Choosing $q_\alpha$ in the interval $q_\alpha \sim (10^6 \div 10^{8})$, a
reheating temperature $T_{RH} \simeq 10^{10}GeV$ and for heavy
right--handed Majorana neutrinos, $\overline{m}
\simeq 10^{15}~GeV$, one gets
\be
\sum_{\alpha=1}^{n_g} \epsilon_{\alpha} \sim \left( 10^{-2} \div 1
\right) \vv
\label{31}
\ee
where the value for the ratio $n_B/s \simeq n_L/s
\sim (10^{-11} \div 10^{-10})$ given by primordial
nucleosynthesis has been used. Note that the lower bound for
$\sum_{\alpha=1}^{n_g}
\epsilon_{\alpha}$ of eq. (\ref{31}) strongly decreases if one
slightly releases the very {\it conservative} condition for no decoherence
(\ref{18bis}). Moreover, since very heavy Majorana neutrinos with mass up
to $10^{18}~GeV$ are still compatible with preheating scenario described in
Ref. \cite{Giudiceetal}, it is interesting to observe that just increasing
the value of $\overline{m}$ of one order of magnitude, by virtue of
(\ref{11-9}), (\ref{18}) and (\ref{30ter}) one reduces the lower bound for
$\sum_{\alpha=1}^{n_g} \epsilon_{\alpha}$ of the same amount.

From the above considerations it follows that the required value for the
ratio $n_B/s$ can be easily obtained for a wide range of the involved
parameters, without imposing fine tuned conditions. In particular it is
worth noticing that no particular mass degeneracy is necessary, and $CP$
violating effects of the same order of those predicted in the electroweak
Standard Model, and measured in $K^0-\overline{K}^0$ system, are already
sufficent to produce a baryon asymmetry of the correct order of magnitude.

\section{Conclusions}

In this paper we have considered a new scenario for the production of a
primordial lepton number, based on decays of oscillating heavy Majorana
neutrinos. In the framework of the preheating mechanism for a non--thermal
production of massive fermions, we have stressed the possible role in
leptogenesis of coherent superpositions of unstable mass eigenstate
neutrinos. The mechanism is similar to the way an observable $CP$ asymmetry
is produced in the neutral $B$ meson system, due to the
$B^0-\overline{B}^0$ oscillation in time. In fact if the decaying neutrinos
are linear superpositions of mass eigenstates $N_i$, the observability of
the $CP$ violation is achieved if the different $N_i$ may interfere.

We have shown that if at least two neutrinos have small mass difference
$\Delta m$, compared with their corresponding decay widths $\Gamma$, a
microscopic asymmetry can be obtained at tree level in the Yukawa matrices,
coupling these states to massless fermions and Higgses. The microscopic
asymmetry is found to be proportional, for small mass differences to the
ratio $\Delta m/\Gamma$. On the other hand, if the neutrino states have
masses quite well separated, with respect to their decay widths, the
interference effects we have described is vanishing and to get a
non--vanishing microscopic asymmetry one should consider interference of
the decay amplitudes at tree level with higher loop contributions
\cite{FukugitaYanagida},\cite{Giudiceetal}--\cite{Pilaftsis}.

In this scenario, a crucial feature is that neutrinos are produced with a
non--thermal distribution, thus the Sacharov out--of--equilibrium condition
is implemented from the very beginning. If, as we have considered,
neutrinos rapidly decay before any thermalization may occur, it is possible
to avoid any wash out of the final lepton number so produced by inverse
processes or scattering, provided the maximal temperature achieved during
reheating is smaller than the mass of the lightest of the heavy Majorana.

It is finally worth stressing that, even using very conservative bounds on
the couplings involved and Majorana masses, the order of magnitude of the
microscopic asymmetries results compatible with $CP$ violation effects in
the Standard Model. Actually this result holds for choices of $\Delta m$
which are not particularly fine tuned.

\noindent
{\large \bf Acknowledgements}\\ One of us, G. Miele, would like to thank A.
Riotto for a useful discussion and valuable comments.

\noindent
{\large \bf Note added in proof}\\ Shortly after completing the paper, the
role of neutrino instability in mixing phenomenon have been presented in a
recent manuscript \cite{Lalakulich}.

\end{document}